\documentclass[12pt]{scrartcl}
\usepackage{geometry}
\usepackage{mathtools}
\usepackage[boxruled,linesnumbered]{algorithm2e}
\usepackage{blindtext, graphicx}
\usepackage{url}
\usepackage{relsize}
\usepackage{cite}
\usepackage[breaklinks,colorlinks=true,citecolor=blue,pdftex]{hyperref}
\usepackage{bm}
\usepackage{changepage}

\begin{document}

\title{DDCCast: Meeting Point to Multipoint Transfer Deadlines Across Datacenters using ALAP\thanks{As Late As Possible \cite{rcd, dcroute}}\hspace{0.4em} Scheduling Policy}

\subtitle{\lbrack Technical Report\rbrack}

\author{\large
    Mohammad Noormohammadpour and Cauligi S. Raghavendra\\ 
    \textit{\large Ming Hsieh Department of Electrical Engineering}\\ 
    \textit{\large University of Southern California (USC)}
}

\date{}

\maketitle

\begin{abstract}
Large cloud companies manage dozens of datacenters across the globe connected using dedicated inter-datacenter networks. An important application of these networks is data replication which is done for purposes such as increased resiliency via making backup copies, getting data closer to users for reduced delay and WAN bandwidth usage, and global load balancing. These replications usually lead to network transfers with deadlines that determine the time prior to which all datacenters should have a copy of the data. Inter-datacenter networks have limited capacity and need be utilized efficiently to maximize performance. In this report, we focus on applications that transfer multiple copies of objects from one datacenter to several datacenters given deadline constraints. Existing solutions are either deadline agnostic, or only consider point-to-point transfers. We propose DDCCast, a simple yet effective deadline aware point to multipoint technique based on DCCast \cite{dccast} and using ALAP traffic allocation \cite{dcroute}. DDCCast performs careful admission control using temporal planning, uses rate-allocation and rate-limiting to avoid congestion and sends traffic over forwarding trees that are carefully selected to reduce bandwidth usage and maximize deadline meet rate. We perform experiments confirming DDCCast's potential to reduce total bandwidth usage by up to $45\%$ while admitting up to $25\%$ more traffic into the network compared to existing solutions that guarantee deadlines.
\end{abstract}

\section{Introduction}
Cloud companies build more and more datacenters globally getting data closer to users and improving users' quality of experience \cite{azure, google, aws}. Focusing datacenters on serving local users saves bandwidth and increases average throughput to users (e.g. TCP's average throughput depends on latency between sender and receiver). This requires replication of data across multiple datacenters and creates many inter-datacenter transfers \cite{orchestrating, tempus, swan, b4}. Replication of data from one datacenter to multiple datacenters results in creation of Point to Multipoint (P2MP) transfers \cite{dccast}. Many such transfers are initiated by non-realtime applications such as backup, data migration and updates \cite{netstitcher}. An overview of various ways P2MP transfers are performed today is provided in \cite{dccast}. 

Many applications require timely completion of transfers which can be modeled as some kind of deadline (hard or soft deadlines) \cite{amoeba, tempus}. Several works aim at guaranteeing deadlines for large inter-datacenter transfers \cite{dtb, mbdt_initial, ecoflow, orchestrating, amoeba, dcroute} by considering a constraint that enforces delivery of all data before a deadline. A deadline may be far enough to allow for flexibility in scheduling transfers, meaning not all transfers need be finished as early as possible. The objective is to maximize the volume of admitted traffic, in some cases considering their priorities \cite{orchestrating}. In addition, deadline-aware transport protocols \cite{d2tcp, d3, pdq} have been developed that target intra-datacenter networks and aim to minimize deadline miss rate for soft real-time applications without admission control.

Deadline transfers considered in prior work take place between specific $(source, destination)$ pairs. If an object is to be delivered to several datacenters while meeting deadlines, multiple point-to-point transfers need to be started and there is no means to plan and guarantee deadlines for P2MP transfers.

In this report, we focus on P2MP transfers with deadlines. We propose \textbf{DDCCast}, a deadline aware point to multipoint solution based on DCCast \cite{dccast} and using ALAP rate-allocation \cite{rcd, dcroute}. DDCCast performs careful admission control using temporal planning \cite{tempus}, forwarding tree selection and rate-allocation \cite{dccast} to reduce bandwidth usage and maximize admitted traffic. We perform experiments confirming DDCCast's potential to reduce total bandwidth usage by up to $45\%$ while admitting up to $25\%$ more traffic into the network compared to existing solutions that guarantee deadlines \cite{amoeba, dcroute}.

\section{Problem Statement and Formulation}
We model each P2MP transfer $R$ as $(T_{D_R}, V_R, S_R, \pmb{\mathrm{D_R}})$, where $T_{D_R}$ determines the deadline prior to which the object with volume $V_R$ has to be delivered from $S_R$ to $n > 0$ datacenters $\pmb{\mathrm{D_R}} = \{D_{1_R},\dots,D_{n_R}\}$. To provide flexible bandwidth allocation, we consider a slotted timeline \cite{tempus, amoeba, dcroute} where the transmission rate of senders is constant during each slot, but can be updated from one slot to the next. This can be achieved using rate-limiting techniques at the end-points \cite{swan, bwe}.

A central scheduler is assumed that receives transfer requests from end-points, performs admission control to determine feasibility, calculates an initial temporal schedule, and informs the end-points of next timeslot's rate-allocation when the timeslot begins. The allocation for future slots can change as new requests are submitted, however, only the scheduler knows about schedules beyond the current timeslot and it can update such schedules as new requests are submitted. We focus on scheduling large transfers that can take minutes or more to complete \cite{tempus} and therefore, the time to submit a transfer request, calculate the routes, and install forwarding rules is considered negligible in comparison. We also assume equal link capacity for all links to simplify the problem. We consider an online scenario where requests may arrive at any time and go through an admission control process; if admitted, they are scheduled to be completed prior to their deadlines. To prevent thrashing, similar to previous works \cite{amoeba, dcroute}, we also assume that once a request is admitted, it cannot be evicted.

A transfer request $R$ is considered \textbf{active} if it has been admitted but not completed. At any moment, there may be $K$ different active requests with various deadlines. We define \textbf{active window} as the range of time from $t_{now}+1$ (next timeslot) to $t_{end}$, the timeslot of the latest deadline, defined as $\max(T_{D_{1}},\dots,T_{D_{K}})$. At the end of each timeslot, all requests can be updated to reflect their remaining (residual) demands by deducting volume sent during a timeslot from their total demand at the beginning of a timeslot. To perform a P2MP transfer $R$, the source $S_R$ transmits traffic over a Steiner Tree \cite{steiner_tree_problem} that spans across all destinations $D_{1_R}$ to $D_{n_R}$ which we refer to as P2MP request's forwarding tree. The transmission rate over a forwarding tree at every timeslot is the minimum of available bandwidth over all edges of the tree at that timeslot. 

\textbf{P2MP Deadline Assumption:} We make the assumption that a P2MP transfer is only valuable if all of its destinations receive the associated object prior to the specified deadline. As a result, a transfer should only be accepted if this requirement can be guaranteed given no failures.

\textbf{P2MP Deadline Problem:} \textit{Determine feasibility of allocating transfer $R_{K+1}$ using some forwarding tree over the inter-datacenter network, given $K$ active requests $R_1$ to $R_K$ with residual demands $V_{R_1}$ to $V_{R_K}$ each with their own forwarding trees. If feasible, determine the forwarding tree that minimizes overall bandwidth usage while taking into account the distribution of load across the network by shifting load to less utilized edges when profitable.}

This problem is along the lines of prior work DCRoute \cite{dcroute} that routes and schedules point to point transfers and DCCast \cite{dccast} which schedules P2MP transfers with focus on decreasing tail times.

\section{General Approach}
The most general approach to solving the \textit{P2MP Deadline Problem} is to form a Mixed Linear Integer Program (MILP) that considers capacity of links over various timeslots along with transfer deadlines and reschedules all active requests along with the new request. The solution would be a new schedule for every active transfer (over the same trees) and a new tree with a rate allocation schedule for new request. Solving MILPs can be computationally intensive and may take a long time. This is especially problematic if MIPs have to be solved upon arrival of requests for admission control (the algorithm needs to check feasibility fast). To speed up this process, we propose DDCCast, which is a heuristic algorithm and is developed as an extension to DCRoute \cite{dcroute} and DCCast \cite{dccast}.

\section{DDCCast}
\begin{figure}
    \centering
    \includegraphics[width=\textwidth]{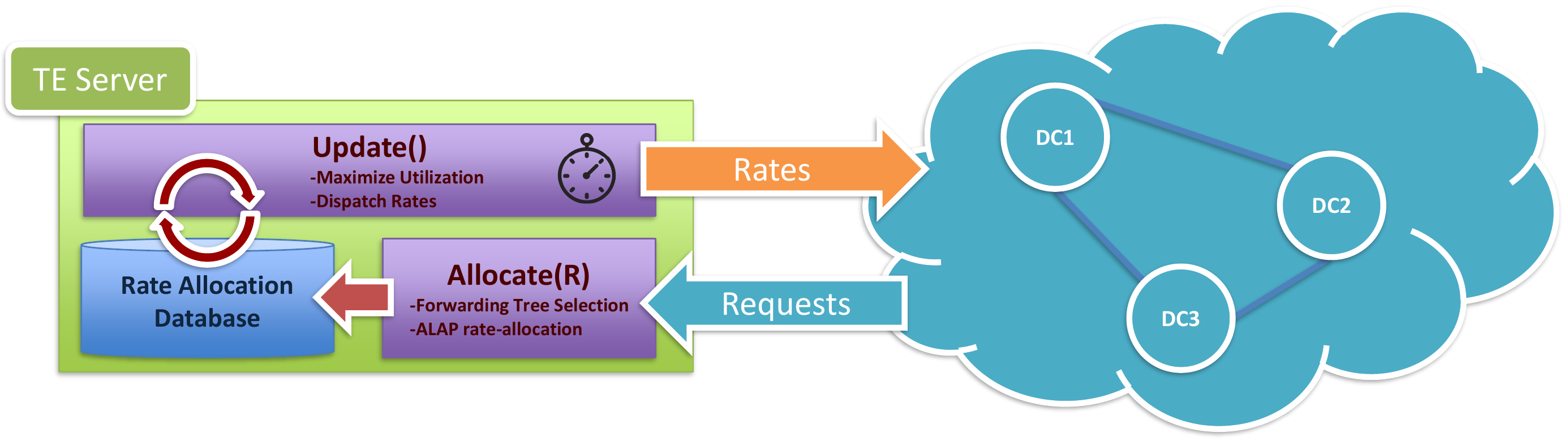}
    \caption{DDCCast (Deadline-Aware DCCast \cite{dccast, dccastgit}) Architecture}
    \label{fig:ddccast}
\end{figure}

The architecture of DDCCast (Deadline-Aware DCCast \cite{dccast, dccastgit}) has been shown in Figure \ref{fig:ddccast}. There are two main procedures of \textbf{Update} and \textbf{Allocate}. The former simply reads the rate-allocations from the database and dispatches them to all end-points at the beginning of every timeslot. The latter performs admission control, forwarding tree selection (in the same way as DCCast) and rate-allocation according to ALAP policy (in the same way as DCRoute \cite{dcroute}). The rates are then updated in a database. Also, at the beginning of every timeslot, if there is unused capacity, the \textbf{Update} procedure moves back some of the future allocations, starting with the closest allocation to the current timeslot that can be moved back, to maximize utilization (the two circular arrows). We will discuss the main parts of DDCCast in the following.

\textbf{Forwarding Tree Selection:} For every new transfer, this procedure selects a forwarding tree that connects the sender to all receivers over the inter-datacenter network. This is done by assigning costs to edges of the inter-datacenter network and selecting a minimum weight Steiner Tree \cite{steiner_tree_problem}. Cost of a forwarding tree is sum of the costs of its edges. For every transfer $R$ with volume $V_R$, we assign edge $e$ of the network the following cost $W_{e,R}$ (also referred to as weight):

\begin{align}
    L_{e,R} &= \sum_{t = t_{now}+1}^{t = T_{D_R}} r_e(t) \\
    W_{e,R} &= V_R + L_{e,R}
\end{align}

Where $t_{now}$ refers to current timeslot, $r_e(t)$ is the total rate allocated on edge $e$ to other requests at time $t$, $L_{e,R}$ calculates the total load on edge $e$ prior to deadline of $R$. Running a minimum weight Steiner Tree heuristic gives us a forwarding tree $T$. This process is performed only once for every request upon their arrival. 

Ideally for routing, we seek a tree with minimum number of edges that connects the source datacenter to all destination datacenters (minimum edge Steiner Tree), but such tree may not have enough capacity available on all edges to complete the transfer prior to its deadline. Therefore, a different Steiner Tree, which can be larger but possesses a higher available bandwidth, may be chosen. It is possible that larger trees provide higher available capacity by using longer paths through least loaded edges, but consume more bandwidth since they send same traffic over a larger number of edges. To model this behavior, we use a cost function that allows balancing two possibly conflicting objectives: finding the forwarding tree with highest available capacity by potentially taking longer paths (to balance load across the network), while minimizing the total network capacity used by minimizing the number of edges used. Our evaluations in \cite{dccast} show that this cost assignment performs more effectively compared to minimizing the maximum utilization in the network which is a well-known policy that is frequently used in traffic engineering literature.

\textbf{Admission Control:} After finding a P2MP forwarding tree, we need to first verify if the new transfer can be accommodated over the tree. We perform the admission control by calculating the available bandwidth over the tree for all timeslots of $t_{now}+1$ to $T_{D_R}$. We then sum the available bandwidth across these timeslots and admit the request if the total is not less than $V_R$. This is shown in following equations:

\begin{align}
    A_T(t) &= \min_{e \in \pmb{\mathrm{E_T}}}(1 - r_e(t)) &\forall t \in \{t_{now}+1,\dots,T_{D_R}\} \\
    A &= \sum_{t = t_{now}+1}^{t = T_{D_R}} A_T(t)
\end{align}

Where $\pmb{\mathrm{E_T}}$ is the set of edges of forwarding tree $T$, $A_T(t)$ is the available bandwidth over $T$ at time $t$ and $A$ is the total available bandwidth over $T$ prior to $R$'s deadline. We accept $R$ if and only if $A \ge V_R$.

\textit{Note:} This admission control approach does not guarantee that a rejected request could not have been accommodated. It is possible that a request is rejected although it could have been accepted if a different forwarding tree had been chosen. In general, finding the tree with maximum available bandwidth prior to a deadline is a complex problem given that maximum available rate over a tree is the minimum of what is available over its edges per timeslot. In addition, even if this problem could be optimally solved, it would still not lead to an overall optimal solution (the overall problem has similarities to online packing problems with multiple capacity constraints).

\textbf{ALAP Rate-Allocation:} Once admitted, rate-allocation process places every new request according to As Late As Possible (ALAP) policy \cite{dcroute} which guarantees meeting deadlines but postpones use of resources until it is absolutely necessary.

\textbf{Rate-Allocation Adjustments:} Adjustments are done in \textbf{Update} procedure upon beginning of timeslots. To maximize utilization, we adjust the schedules when there is unused capacity. Upon beginning of every timeslot, we pull traffic from closest timeslots in the future over each P2MP forwarding tree and send it in current timeslot, if there is available capacity along all edges of such a P2MP forwarding tree. For a network, it may not be possible to schedule traffic ALAP on all edges since allocations may need to span over multiple edges all of which may not have available bandwidth. Therefore, after maximizing utilization by pulling traffic to current timeslot, we scan the timeline starting the next future timeslot and push allocations forward as much as possible until no schedule can be pushed further toward its deadline.

\section{Evaluation}
We evaluated DDCCast using synthetic traffic generated in accordance with earlier works \cite{amoeba, dcroute, rcd, dccast}. The arrival of requests followed a Poisson distribution with rate $\lambda$. The deadline $T_{D_R}$ of every request $R$ was generated using an exponential distribution with mean of $10$ timeslots. Demand of $R$ was then calculated using another exponential distribution with mean $\frac{T_{D_R}-t_{now}}{8}$. All simulations were performed over $500$ timeslots and each scenario was repeated $10$ times and the average measurements have been reported. We assumed a total capacity of $1.0$ for each timeslot over each link.

We performed our simulations over Google's GScale topology \cite{b4} with $12$ datacenters and $19$ links. We assumed a machine attached to each datacenter generating traffic destined to other (multiple) datacenters. The simulations were performed on a single machine (Intel Core i7-6700T CPU and 24GBs of RAM). All simulations were coded in Java, and to solve linear programs for Amoeba, we used Gurobi Optimizer \cite{gurobi}. Please note that DDCCast behaves almost identical to DCRoute for one-to-one transfers and identical to DCCast when deadlines approach infinity.

We measured two metrics: \textbf{total bandwidth used} and \textbf{total traffic admitted}. Both parameters were calculated over the whole network and all timeslots. The first parameter is the sum of all traffic over all timeslots and all links. The second parameter determines what volume of offered load from all end-points was admitted into network.

Following schemes were considered: DDCCast, DCRoute \cite{dcroute} and Amoeba \cite{amoeba} all of which aim to guarantee the deadlines, maximize total utilization, and perform initial admission control. DCRoute and Amoeba do not have the notion of point to multipoint forwarding trees. As a result, to perform the following simulations, each P2MP transfer with multiple destinations in DCCast is broken into several independent point-to-point transfers from source to each destination and then plugged into DCRoute and Amoeba.

We only compare DDCCast with these two works since other works either do not support deadlines \cite{bwe, swan} or focus on different objectives. For example, some aim to maximize the fraction of transfers completed prior to deadlines applying fair sharing policy \cite{tempus} or to minimize total delivery costs incurred by minimizing the maximum utilization over ISP provided infrastructure (by leveraging diurnal patterns) \cite{dtb, netstitcher, ecoflow, mbdt_initial, jetway}. We also do not compare with techniques based on store-and-forward such as \cite{orchestrating}. These schemes may incur excessive delays in delivering large transfers, as a function of how many times they store each object on intermediate datacenters, which could be problematic to transfers with tight deadlines. They also require storing multiple copies of objects on intermediate nodes incurring storage costs as the number of transfers increase, and wasting intra-datacenter bandwidth since such objects have to be stored on a server inside the intermediate datacenters.

\subsection{Effect of number of destinations}
\begin{figure}
    \centering
    \includegraphics[width=\textwidth]{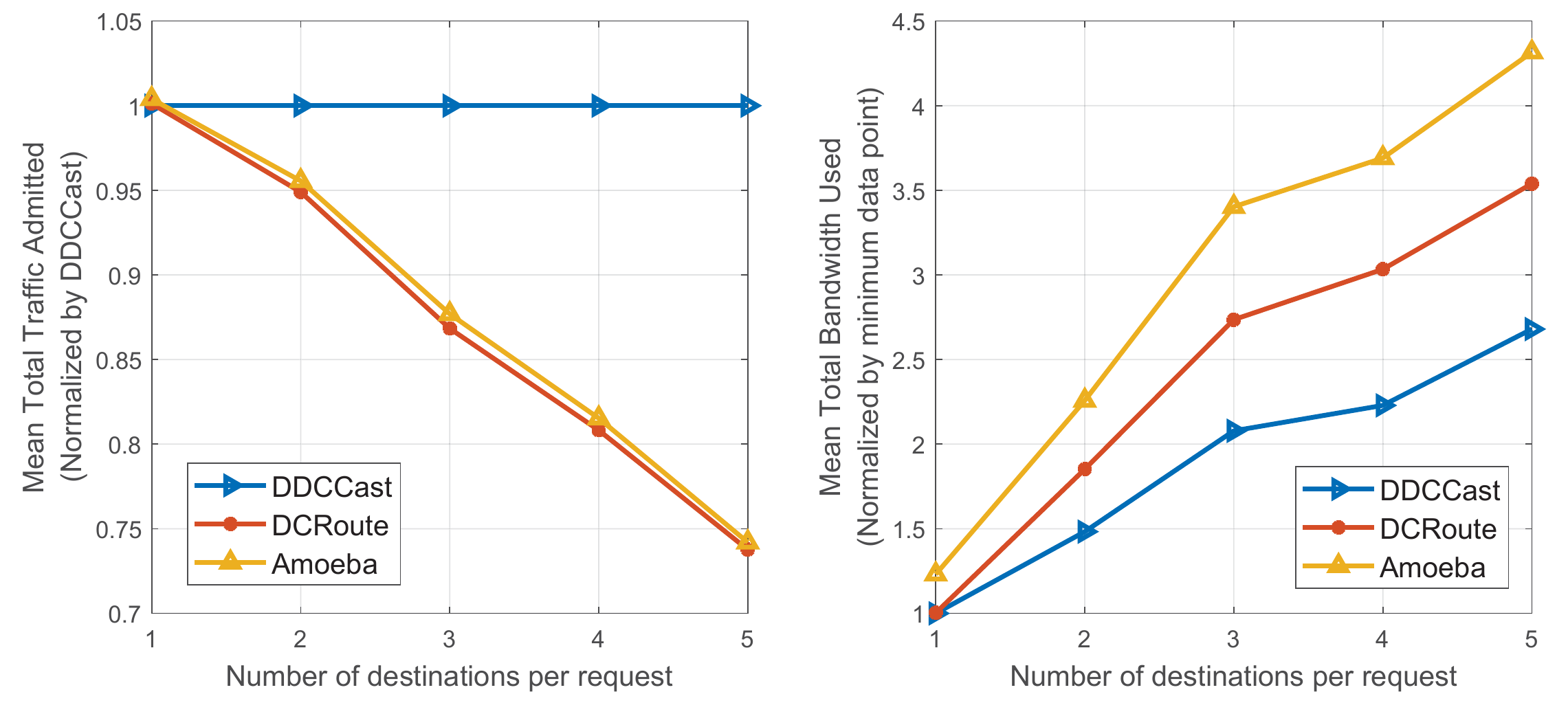}
    \caption{Bandwidth Usage and Admitted Traffic vs. the Count of Destinations ($\lambda_{P2MP} = 2$)}
    \label{fig:p2mp_copies}
\end{figure}

Figure \ref{fig:p2mp_copies} shows the results of this experiment. We increased the number of destinations for each transfer from $1$ to $5$ and picked random destinations for each transfer. The total volume of traffic used by Amoeba \cite{amoeba} is up to $1.8$ times the volume used by DDCCast. Even in case of one destination Amoeba uses $1.2$ times the bandwidth of DCCast and DCRoute. This occurs because Amoeba routes traffic across $K$-shortest paths some of which may not be as short as others. Therefore, even for small loads, a portion of traffic may traverse longer paths and increase total bandwidth usage. 

DDCCast saves bandwidth by using P2MP forwarding trees and using a single path to each destination which is chosen to be as short as possible while avoiding highly loaded edges. This way, DDCCast admits $25\%$ more traffic compared to Amoeba when sending objects to $5$ destinations while using $45\%$ less overall network capacity.

\subsection{Effect of request arrival rate (load)}
\begin{figure}
    \centering
    \includegraphics[width=\textwidth]{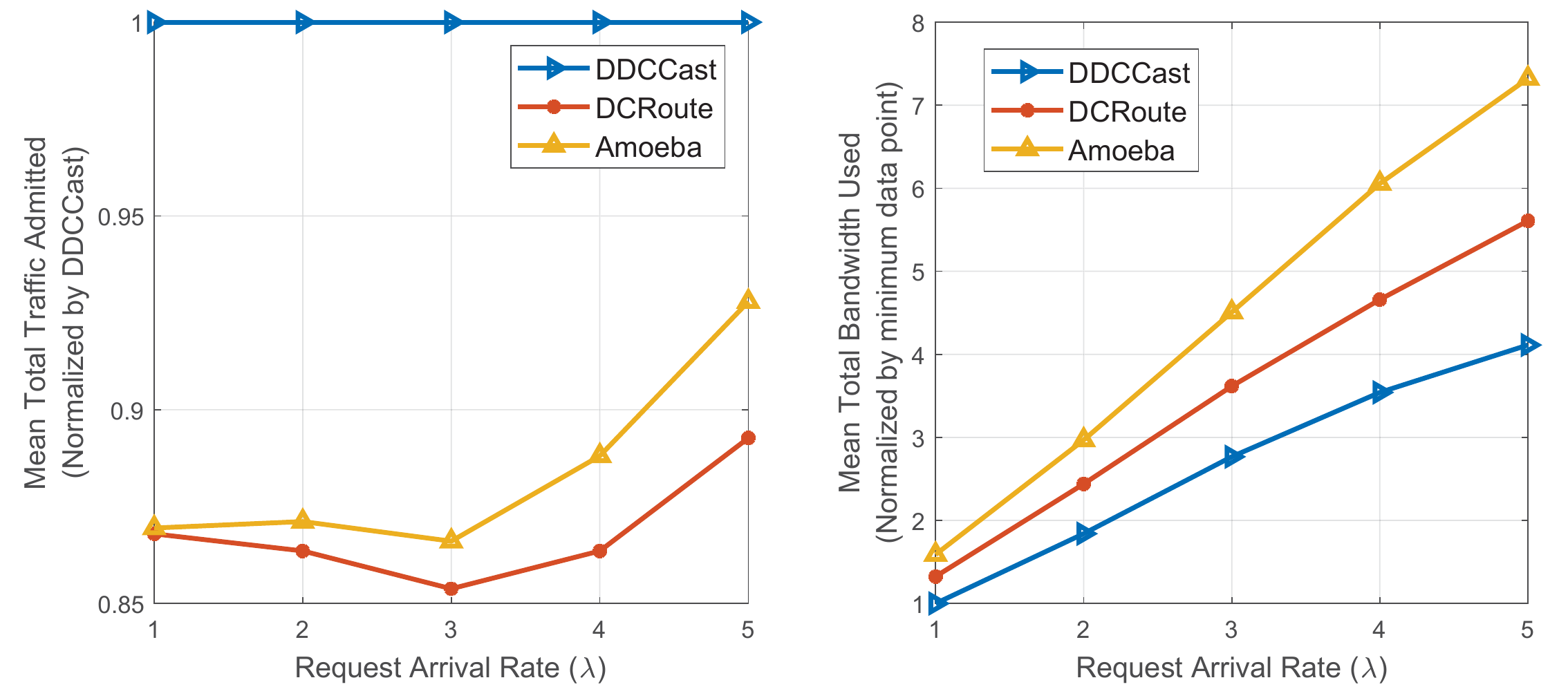}
    \caption{Bandwidth Usage and Admitted Traffic vs. the Request Arrival Rate ($\#~ Destinations = 3$)}
    \label{fig:p2mp_lambda}
\end{figure}

In this experiment, we investigate the effect of $\lambda$ while sending an object to three destinations. Results of this experiment have been shown in Figure \ref{fig:p2mp_lambda}. Volume of admitted traffic is about $10\%$ higher for DDCCast compared with other two schemes over all arrival rates. Also, similar to previous experiment, DDCCast's total bandwidth usage is between $37\%$ to $45\%$ less than Amoeba \cite{amoeba} and $28\%$ less than DCRoute \cite{dcroute}.

\subsection{Computational Overhead of DDCCast}
The computational speed of DDCCast depends on heuristic used for calculation of Steiner trees. Such heuristics are usually fast and often provide close to optimal solutions. In addition, the rate-adjustment applied in DDCCast is similar to techniques used in DCRoute which were shown to impose inconsiderable overhead.

\section{Conclusions}
In this report, we presented DDCCast, which aims to reduce the total network bandwidth usage while guaranteeing deadlines for point to multipoint transfers. DDCCast relies on creation of on demand forwarding trees which we refer to as P2MP forwarding trees. It is possible to create such trees using commodity hardware, SDN frameworks such as OpenFlow \cite{openflow}, and application of Group Tables \cite{openflow-1.3.1} with increasing support in switches \cite{of-juniper-explain, of-juniper, of-huawei, of-hp, of-hp-2}. Such trees can be configured upon arrival of transfers and torn down upon their completion. Our evaluations show that DDCCast admits more load into the network while using less total capacity, over a variety of request arrival rates and number of destinations for requests.

{\footnotesize \bibliographystyle{unsrt}
\bibliography{citations}}

\end{document}